\documentclass{article}
\usepackage[utf8]{inputenc}
\usepackage{amsmath}
\usepackage{amsthm}
\usepackage{amsfonts}
\usepackage{amssymb}
\usepackage{graphicx}
\usepackage{float}
\usepackage{caption}
\usepackage{subcaption}
\usepackage{textcomp}
\usepackage{mathtools}
\usepackage{physics}
\usepackage{comment}

\usepackage{authblk}
\usepackage{enumitem}

\usepackage{hyperref}

\usepackage{xcolor}

\usepackage[printonlyused]{acronym}

\providecommand{\keywords}[1]
{
  \small	
  \textbf{\textit{Keywords:}} #1
}

\DeclarePairedDelimiter\floor{\lfloor}{\rfloor}

\title{Optical payload design for downlink quantum key distribution and keyless communication using CubeSats}

\author[1,2]{Pedro Neto Mendes}
\author[2,3]{Gonçalo Lobato Teixeira}
\author[1]{David Pinho}
\author[2]{Rui Rocha}
\author[1,2]{Paulo André}
\author[2,4]{Manfred Niehus}
\author[2]{Ricardo Faleiro}
\author[5,6,7]{Davide Rusca}
\author[1,2]{Emmanuel Zambrini Cruzeiro\footnote{emmanuel.cruzeiro@lx.it.pt}}
\affil[1]{Departamento de Engenharia Electrotécnica e de Computadores, Instituto Superior Técnico, Av. Rovisco Pais, 1049-001, Lisbon, Portugal}
\affil[2]{Instituto de Telecomunicações, Av. Rovisco Pais, 1049-001, Lisbon, Portugal}
\affil[3]{Departamento de Física, Instituto Superior Técnico, Av. Rovisco Pais, 1049-001, Lisbon, Portugal}
\affil[4]{Instituto Superior de Engenharia de Lisboa, R. Conselheiro Emídio Navarro 1, 1959-007, Lisbon, Portugal}
\affil[5]{Vigo Quantum Communication Center, University of Vigo, Vigo E-36310, Spain}
\affil[6]{Escuela de Ingeniería de Telecomunicación, Department of Signal Theory and Communications, University of Vigo, Vigo E-36310, Spain}
\affil[7]{AtlanTTic Research Center, University of Vigo, Vigo E-36310, Spain}

\begin{document}

\maketitle

\begin{abstract}
Quantum key distribution is costly and, at the moment, offers low performance in space applications. Other more recent protocols could offer a potential practical solution to this problem. In this work, a preliminary optical payload design using commercial off-the-shelf elements for a quantum communication downlink in a 3U CubeSat is proposed. It is shown that this quantum state emitter allows the establishment of two types of quantum communication between the satellite and the ground station: quantum key distribution and quantum keyless private communication. Numerical simulations are provided that show the feasibility of the scheme for both protocols as well as their performance. For the simplified BB84, a maximum secret key rate of about 80 kHz and minimum QBER of slightly more than $0.07\ \%$ is found, at the zenith, while for quantum private keyless communication, a 700 MHz private rate is achieved. This design serves as a platform for the implementation of novel quantum communication protocols that can improve the performance of quantum communications in space. 
\end{abstract}


\section{Introduction}

\subsection{Long distance quantum communication}

In quantum communication, physical systems are exploited to encode and transfer information between parties. Thanks to C. Shannon \cite{Shannon1948} and to the second quantum revolution, physicists began to develop a new understanding of what information is. This has led to newly emerging technological applications, such as quantum communication, quantum computation, quantum sensing, and quantum thermodynamics \cite{chugh2023progression, crawford2021quantum}.

Quantum communication promises unconditional security based on the laws of nature without needing to impose requirements on the computational power available to an eavesdropper, which might, at first sight, seem surprising. The most celebrated variant of quantum communication is \ac{QKD}, which is proven to achieve, under certain assumptions, such unconditional security. The most impressive demonstrations of \ac{QKD} were implementations of so-called device-independent \ac{QKD} protocols, which allow unconditional security with no assumptions about the inner workings of the devices used to distribute the key. These demonstrations were only performed very recently \cite{Nadlinger2021,Zhang2021,Liu2021}, almost 40 years after the proposal of the first \ac{QKD} protocol, the BB84 \cite{Bennett1983}.

\ac{QKD} has seen considerable progress in the last decade, as illustrated by the development of commercial systems \cite{comercial} since the early 2000s. Nevertheless, the available devices carry a glaring limitation: the rate-distance trade-off. Even with low-loss fibers, commercial \ac{QKD} systems are limited to a few hundred kilometers for a useful key rate \cite{Zhang:18, continuous}. Therefore, in terms of long-distance telecommunications, \ac{QKD} is still in its infancy \cite{Pirandola_2020}. There are two main approaches to extending \ac{QKD} to distances of hundreds to thousands of km: quantum repeaters and space-based \ac{QKD}. This work focuses on the latter, by implementing the simplified BB84 protocol \cite{Grunenfelder2018} between a nanosatellite in \ac{LEO} and a ground station.

Although it is the most popular form of quantum communication, the assumptions behind the security proofs of \ac{QKD} are very strong, as they consider a wide generality of possible attacks by a malicious agent. In fact, these assumptions may be unnecessarily demanding for satellite-to-ground station communication, due to physical limitations on Eve's ability to completely intercept and resend information without being detected. A more reasonable solution, in this case, is quantum keyless communication \cite{Vazquez2021}, whereby information is directly sent over the quantum channel, encoded in the quantum states of light. There is no key generation in this case. Therefore a design that can serve as a quantum state emitter both for \ac{QKD}, and to implement \ac{QKPC} is proposed.

In this work, the initial design of an optical payload for a 3U CubeSat downlink is described. The optical payload consists of a source of quantum states which may be used to both implement the simplified BB84 and \ac{QKPC}. To this end, a compact version of the usual simplified BB84 setup \cite{Grunenfelder2018}, adapted to fit in the restricted volume and power budget of the nanosatellite, is designed. An implementation of the preliminary design is proposed, taking into account optical, mechanical, and electrical design, along with celestial mechanics considerations, and realistic simulations of both protocols are provided. 

This proposal is innovative compared to other proposals for nanosatellite quantum communication for its versatility: it may implement various novel quantum communication protocols, which is demonstrated by its ability to implement \ac{QKD} and \ac{QKPC}. In other words, this solution serves as a starting point for future research in novel quantum communication protocols for space-based applications. The main purpose of this article is to propose a platform for satellite quantum communication experiments beyond quantum key distribution. Additionally, it is shown that the quantum state emitter can fit inside a 3U CubeSat, using only commercial off-the-shelf elements.

\subsection{Satellites for quantum communication - overview}

In terms of satellite communication, \ac{QKD} is still in its infancy \cite{bed2016,polnik2020}. In 2003, an experiment in the Matera Laser Ranging Observatory (Italy) demonstrated the feasibility of sending single photons through the atmosphere in a ground-\ac{LEO}-ground link \cite{villoresi2008experimental}. This showed that a global \ac{QKD} network may indeed be created in the future with a mix of satellite and ground nodes. Japan and China both created road maps to develop this technology which led to the launch of SOCRATES \cite{takenaka2017satellite} and Micius \cite{Lu_2022} in 2014 and 2016 respectively.

SOCRATES is a Japanese micro-satellite in \ac{LEO} orbit, weighing 48 kg, measuring 496 x 495 x 485 mm, and whose goal is to establish a standard micro-satellite bus technology applicable to missions of various purposes. Inside it, SOTA (Small Optical TrAnsponder), the small and light (6 kg weight, 17.8 x 11.4 x 26.8 cm) optical quantum-communication transmitter, allowed to perform various experiments that culminated in \ac{LEO}-to-ground quantum communication in 2017 \cite{takenaka2017satellite}.

Micius is a Chinese satellite in \ac{LEO} orbit, weighing 635 kg part of QUESS, a proof-of-concept mission designed to facilitate quantum optics experiments over long distances to allow the development of quantum encryption and quantum teleportation technology. The satellite consists of two transmitters. Transmitter 1, weighing 115 kg, incorporates eight laser diodes and a BB84 coding module to facilitate \ac{QKD} through preparation and measurement. The second transmitter, weighing 83 kg, is specifically designed to distribute quantum entanglement from the satellite to two distinct ground stations. Within a year of the launch, three key milestones for a global-scale quantum communication network were achieved: satellite-to-ground decoy-state \ac{QKD} with KHz rate over a distance of up to 1200 km;  satellite-based entanglement distribution to two locations on Earth separated by $\approx$ 1200 km and the subsequent Bell test, allowing possible effective link efficiencies through
satellite of 12-20 orders of magnitudes greater than direct transmission; ground-to-satellite quantum teleportation \cite{Lu_2022}.

\subsection{Quantum CubeSats state of the art}

Recently, the focus on space-based quantum communication shifted to smaller satellites, specifically CubeSats, which are the most common type of nanosatellite. In the last decade, the use of CubeSats has grown considerably \cite{villela2019towards}. These systems are interesting because they are cost-effective, are easier and faster to develop, and can ride along in rockets designed for different payloads. This has allowed companies, non-profit organizations, and even educational institutions to participate in their development and launch.

This contributed to the creation of various research projects to develop CubeSats for quantum communication all around the world. These projects started with path-finders works like CQuCoM \cite{Oi2017}, followed by specific missions. Germany started the QUBE project \cite{haber2018qube} to develop a 3U CubeSat for downlink \ac{QKD} implementation. In France, the Grenoble University Space Center is leading the development of NanoBob, a 12U CubeSat to demonstrate the feasibility of quantum communication over a distance of 500 km. NanoBob \cite{de2022satellite, nanobob} is expected to launch in 2024 and Grenoble University Space Center is already engaged in a more ambitious project, financed by the French Space Agency CNES and with TAS-F as the leading partner, that investigates the requirements and specification of a future Quantum Information Network that includes one or more Space links. Companies are also collaborating with academia in satellite-based \ac{QKD} projects, like ROKS mission \cite{zhang2023}, a 6U CubeSat with a 1/3U size optical module employing a 4-state BB84 with \ac{WCP}, set to launch in 2024. Other missions include QEYSSat \cite{jennewein2023qeyssat} and QUARC \cite{mazzarella2020}, aiming to demonstrate the feasibility of quantum links in uplink and downlink configuration.  

For now, efforts for space-based quantum communication have focused mostly on \ac{LEO} orbits. This is because of the relative ease of reaching the orbit, the possibility to cover the entire planet in a matter of hours with a single satellite (rapid round trip and many orbit inclination options), and the more relaxed link budget making it easier to develop a communication system. Nevertheless, this type of orbit has its limits as the passage over a ground terminal is limited to just a few minutes of effective link (lower communication window) and the tracking system has to be more precise. Recently, the first experimental single-photon exchange with a \ac{MEO} satellite at 7000 km was realized \cite{Dequal_2016} followed by a feasibility study for quantum communication at \ac{GEO} orbits (allowing 24-hour link coverage) \cite{gunthner2017quantum}.

These approaches to quantum communication in space focus on \ac{QKD} and, to our knowledge no other quantum communication protocols have been proposed.

\clearpage

\section{Concept and implementation}\label{sec:qkd}

A versatile CubeSat design that allows for various types of quantum communication schemes is proposed. In this section, two protocols that can be implemented with the design are described. The first protocol is a recent variant of the BB84 protocol, called the simplified BB84 \cite{Grunenfelder2018}. The second is a \ac{QKPC} scheme for keyless secure communications \cite{Vazquez2021}.
Then, the setup realizing the protocols is described, and a Size, Weight, and Power analysis of the preliminary design is conducted to validate it.

\subsection{Protocols}

In polarization-based BB84, Alice sends a number of states picked from the following qubit basis,
\begin{align} 
    & |\mathrm{H}\rangle,\ |\mathrm{V}\rangle,\nonumber\\
    & |\mathrm{D}\rangle\coloneqq \frac{1}{\sqrt{2}}(|H\rangle + |V\rangle),\ |\mathrm{A}\rangle\coloneqq \frac{1}{\sqrt{2}}(|H\rangle - |V\rangle), \nonumber\\
    & |R\rangle \coloneqq \frac{1}{\sqrt{2}}(|H\rangle + i|V\rangle),\ |L\rangle \coloneqq \frac{1}{\sqrt{2}}(|H\rangle - i|V\rangle) \label{eq:RL},
\end{align}

where $|.\rangle$ are the polarization states.
Taking advantage of a different subset of the above states, several variants of BB84 exist. The original BB84 used four states, and a more noise-robust version exists with six states, the so-called six-state BB84 protocol \cite{bruss1998,bechmann1999}. Moreover there exists variants which uses only three states (two for the computational basis and one for the monitoring basis) which keep the secret key rate almost unchanged with respect to the original BB84 but allow for a simpler implementation \cite{molotkov1996,shi2000,fung2006,tamaki2014,Rusca2018b}

A naive implementation of BB84 using \acp{WCP} is not secure due to the photon number splitting attack \cite{brassard2000,lutkenhaus2000}. To mitigate this problem, one uses decoy states, i.e. states with different intensities which allow the users to determine more easily the presence of an eavesdropper \cite{hwang2003,lo2005,wang2005,ma2005,hayashi2014,lim2014}. In \cite{Rusca2018,Rusca2020}, a comparison was made between BB84 protocols taking advantage of decoy states. Following these works we found that the best protocol in terms of security and experimental simplicity for our purpose is the simplified BB84 protocol, which uses three states and one decoy and allows for a simpler receiver scheme.

A version of the simplified BB84 protocol with one decoy was first implemented in polarization in \cite{Grunenfelder2018}, in the following, the idea of the protocol is summarized. For the computational basis Z, the protocol runs exactly as the original BB84. However, in the monitoring basis X, Alice prepares only $|D\rangle$, while Bob's measurement corresponds to a projection onto $|A\rangle$. In this protocol, only three preparations and three detections are necessary. The detections can be implemented with two detectors, as in \cite{Grunenfelder2018}.

The protocol is similar to the original BB84,
\begin{enumerate}
    \item \textbf{State preparation}: random encoding in bases X and Z with respective probabilities $p_X^A$ and $p_Z^A = 1 - p_X^A$. In the Z basis, Alice emits $|H\rangle$ and $|V\rangle$ uniformly, while in the X basis she always emits $|D\rangle$. The mean photon number of the pulses is chosen randomly between two values $\mu_1$ and $\mu_2$ with probabilities $p_{\mu_1}, p_{\mu_2}$.\\
    \item \textbf{Measurement}: Bob performs measurements X and Z with respective probabilities $p_X^B$ and $p_Z^B = 1-p_X^B$. He records each basis and measurement outcome.\\
    \item \textbf{Basis reconciliation}: Alice and Bob announce their basis choices for each detection event. Events from the Z basis are used to generate the raw key, while those from the X basis are used to estimate the eavesdropper's potential information. After collecting a number of $n_Z$ raw key bits, they proceed to the next step.\\
    \item \textbf{Error correction/Information reconciliation}: Alice and Bob employ an error correction algorithm on their block of $n_Z$ bits, during which $\lambda = f\cdot n_Z\cdot h(Q_Z)$ bits are disclosed, where $f$ is the reconciliation efficiency, $h(x)$ the binary entropy, and $Q_Z$ the error rate. The procedure succeeds with probability $1-\epsilon_\text{corr}$. After $k = n_Z^*/n_Z$, where both $n_Z^*$ and $n_Z$ are chosen by the users, they proceed to the final step.\\
    \item \textbf{Privacy amplification}: Alice and Bob apply privacy amplification on a block of size $n_Z^*$ to obtain a secret key of $l$ bits (\ac{SKL}), where
    \begin{equation}
        l = \floor*{s_{Z,0}+s_{Z,1}[1-h(\phi_Z)]-\lambda_\text{EC}-6\log_2\left(\frac{\alpha}{\epsilon_s}\right)-\log_2\left(\frac{2}{\epsilon_c}\right)},
    \end{equation}
    where $s_{Z,0}$ is the number of vacuum events, $s_{Z,1}$ is the number of single photon events, and $\phi_Z$ is the phase error rate in the sifted $Z$ basis. $\epsilon_c$ and $\epsilon_c$ are prescribed security parameters, the correctness, and secrecy of the key, respectively. $\lambda_\text{EC}$ is an estimate of the number of bitsrevealed during the error correction. $\alpha = 19 (21)$ for one (two) decoy(s). Finally, $h(\cdot)$ is the binary entropy function. 
\end{enumerate}

The numbers of vacuum and single-photon events $s_{Z,0}$, $s_{Z,1}$ and the phase error rate $\phi_Z$ can be evaluated as described in the SatQuMa documentation \cite{Sidhu2021}. For the evaluation of $\lambda_\text{EC}$, method 1 is used. Statistical fluctuations are evaluated using the Chernoff bound.

The \ac{SKR} is simply the \ac{SKL} divided by the duration of the transmission $T_\text{trans.}$,
\begin{equation}
    \text{SKR} = \frac{\text{SKL}}{T_\text{trans.}}
\end{equation}

For the \ac{QBER} of the X basis, \cite{Rusca2018b},  
\begin{align}
    Q_X & = \frac{1}{2} \frac{P_Z^A P_Z^B}{n_Z} \left[ \frac{n\left(A,D\right)}{P_X^A P_X^B} + \right. \nonumber\\
    & \left. \max \left( 0 , \frac{n\left(A,D\right)}{P_X^A P_X^B} + \frac{n\left(A,Z\right)}{P_Z^A P_X^B} - \frac{n\left(Z,D\right)}{P_X^A P_Z^B} + 2\frac{n_Z}{P_Z^A P_X^B} \right) \right]
\end{align}

is used, where $P_{X(Z)}^{A(B)}$ is the probability of Alice (Bob) sending (measuring) a bit in the $X$ ($Z$) basis. $n_Z$ is the total number of detected bits in the $Z$ basis and the $n(b,a)$ is the number of detections when Alice sends state $a$ and Bob measures state $b$.

The \ac{QKPC} protocol, proposed in \cite{Vazquez2021}, is based on the classic wiretap model, first proposed by Shannon in 1949 \cite{Shannon1949} and later rigorously defined by Wyner in 1975 \cite{Wyner1975} where the author introduced the concept of secret capacity (maximum communication rate at which legitimate users can communicate securely in the presence of an eavesdropper). In the wiretap model, Alice wants to send a message to Bob over a communication channel but a wiretapper (Eve) is listening to the channel. The goal is to encode the data in such a way that maximizes the wiretapper's confusion making it impossible for her to recover the message sent. 

The \ac{QKPC} protocol consists of the following steps:

\begin{enumerate}
    \item \textbf{Encoding}: Alice selects a $n$-bit codeword $X^n$ for her secret message $M$. The secrecy depends on the encoder, which is characterized by the rate $R=k/n$, where $k$ is the number of secret bits, the error probability $\epsilon_n$, and the information leakage measured by an information-theoretical measure denoted $\delta_n$.\\
    \item \textbf{State preparation}: Alice prepares a coherent state modulated by the random variable $X\in \{0,1\}$, where $X=0$ with probability $q$. The \ac{OOK} states are the vacuum state $|\alpha_0\rangle$ and a weak coherent state
    \begin{equation}
        |\alpha_1\rangle = e^{-|\alpha_1|^2/2}\sum_{n=0}^\infty\frac{\alpha_1^n}{(n!)^{1/2}}|n\rangle
    \end{equation}
    The probability $q$ needs to be optimized depending on the assumptions at the detection and the physical propagation channel.\\
    \item \textbf{Measurement}: After $n$ transmissions, Bob receives $B^n$ and Eve $E^n$. Bob obtains $Y^n$ by estimating his received coherent state. Eve can use the best quantum detection strategy to obtain $Z^n$.\\
    \item \textbf{Decoding}: Bob and Eve send their estimated received states to the decoder.
\end{enumerate}

The choices of encoder and decoder are assumed to be public. The values of $\epsilon_n$ and $\delta_n$ depend on these choices.

According to wiretap theory, even if the eavesdropper is computationally unbounded, then
\begin{equation}
    \lim_{n\longrightarrow\infty}\epsilon_n = \lim_{n\longrightarrow\infty}\delta_n = 0
\end{equation}
as long as $R$ is an achievable rate. This means the error probability and information leakage towards Eve can be made arbitrarily low. See \cite{Cai2004,Devetak2005} for exact definitions of the parameters $\epsilon_n$ and $\delta_n$.

\subsection{QKPC protocol security}

For the \ac{QKPC} protocol, when considering satellite and ground station space links some physically motivated limitations on Eve's power can be naturally assumed, like the impracticality of a successful and unnoticed intercept and resend attack over free space. Under such limitations, an eavesdropping attempt can be assumed to exist only for a fraction of the communicated signal, implying that the model used to prove security against Eve may be relaxed, say to a quantum wire-tap model. This in turn opens the door to physical-layer security as a legitimate alternative to \ac{QKD} for establishing secure satellite-to-ground quantum communication \cite{Hayashi2020,Vazquez2021,Ghalaii2022}. \ac{QKPC} allows for much higher rates than \ac{QKD} with current technology \cite{Vazquez2021} and even allows daylight operation, which is presently impractical for \ac{QKD}.

A single-mode free-space quantum bosonic channel is assumed, following \cite{Vazquez2021}. The efficiency of the channel is $\eta$. The channel degradation is described by a parameter $\gamma\in (0,1)$. Therefore the efficiency of Eve's channel is $\gamma\eta$. Bob is assumed to have a single photon detector with limited efficiency (included in $\eta$) and dark count probability $p_\text{dark}$. The stray light is modeled as a Poisson photon number distribution with average $\eta_0\Delta$, where $\eta_0$ is the optical loss between the telescope input lens of the receiver and the detector, and $\Delta$ is the average number of noise photons for a given collection angle and the given frequency and time window, see Appendix D in \cite{Vazquez2021}.

The conditional probabilities of Bob detecting $y$ given that Alice has sent $x$ are given by 
\begin{equation}
    \epsilon_0 = (1-p_\text{dark})e^{-\eta_0\Delta},\quad \epsilon_1 = (1-p_\text{dark})e^{-(\eta\mu+\eta_0\Delta)},
\end{equation}
where $\mu = |\alpha_1|^2$.

Eve is assumed to perform an optimal quantum detection, which leads to an optimal error probability $\epsilon^*$, given explicitly by
\begin{equation}
    \epsilon^*(\gamma) = \frac{1-\sqrt{1-4q(1-q)e^{-\eta\gamma\mu}}}{2}
\end{equation}

The private capacity of \ac{OOK} is then
\begin{equation}
    C_P(\gamma) = \Big[h(\epsilon^*(\gamma)) + h(\frac{\epsilon_0+\epsilon_1}{2})-\frac{h(\epsilon_1)+h(\epsilon_0)}{2}-1\Big]_+
\end{equation}
where $[]_+$ is the positive part and $h(\cdot)$ is the binary Shannon entropy.

Finally, the Devetak-Winter rate for this protocol is given by
\begin{equation}
    R_\text{DW}(\gamma) = I(X,Y) - \chi (X;E|\gamma),
\end{equation}
where $I(X;Y)$ is the Shannon mutual information of Alice's choice of the input probability, measured by a photon counting detector. $\chi (X;E|\gamma)$ is the Holevo bound for Eve, see \cite{Vazquez2021}.

Finally, the rate reduces to
\begin{equation}
    R_\text{DW} = \Big[h(\frac{\epsilon_0+\epsilon_1}{2})-\frac{h(\epsilon_1)+h(\epsilon_0)}{2}-h(\frac{1+\epsilon (\gamma)}{2})\Big]_+
\end{equation}

\subsection{Experimental concept}

\begin{figure}[H]
  \centering
  \includegraphics[width=1\linewidth]{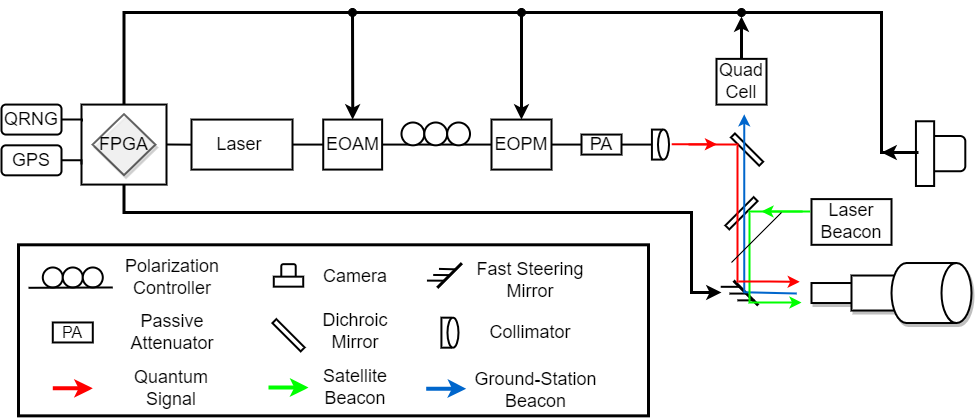}
  \caption{Setup capable of implementing both the \ac{QKPC} and the Simplified BB84 in a 3U Cubesat.}
  \label{fig:ActiveSetupNano}
\end{figure}

The satellite is controlled by an onboard computer that manages the satellite systems (payload, power, etc.), handles data storage and communication, and monitors the health status of the satellite. This system is represented in Fig. \ref{fig:ActiveSetupNano} as an \ac{FPGA} and while a detailed study on how to optimize the on-board computer will be left for future work, an initial estimation of its parameters is used based on information from \cite{Neumann2018}, as it has a similar system. Another solution can be found in \cite{nanobob} by using a commercial Zync-based on-board computer.

A \ac{QRNG} is used to supply a random seed for the choice of basis. The IDQ20MC1 (\ac{QRNG} chip for space applications) from ID Quantique meets all the requirements making it a viable option. Additionally, a GPS module, ACC-GPS-NANO from Accord, is used for time stamping and, as it ensures an accurate determination of orbital position and time, it is also used for the coarse step of the pointing system.

The setup includes a gain-switched distributed feedback (\ac{DFB}) laser source from Anritsu, specifically the \ac{DFB} 1550. Gain-switched lasers are essential to ensure phase randomization of the initial light pulses, as referenced in studies by \cite{Yuan2014} and \cite{Lovic2021}. This source provides coherent phase-randomized pulses at 1550 nm with a pulse duration of 93 ps, triggered at 1 GHz with mW of power. This wavelength is chosen for its easy integration into a fiber-based quantum network, availability of off-the-shelf components due to terrestrial-fiber developments, and high transmittance in the atmosphere \cite{article_1550nm_transmittance}.

The laser is directly coupled into an electro-optic amplitude modulator (\ac{EOAM}), specifically the LN81S-FC from Thorlabs, to encode decoy states via amplitude modulation. A variable waveplate from Phoenixphotonic (\ac{PC}) prepares the state polarization, allowing it to be rotated to any of the three linearly polarized states required for the simplified BB84 protocol. This is achieved by a polarization switch, the PSW-LN-0.1-P-P-FA-FA from IxBlue. Finally, the pulses are attenuated by a passive attenuator, the FA25T from Thorlabs, and exit the fiber through a collimator, the RC04APC-P01 from Thorlabs.

Compared to Grunenfelder \textit{et al.}, this setup was simplified by removing the polarization controller and the high-birefringence fiber after the \ac{EOPM}, as they can be delegated to the ground station, and instead of a variable attenuator, a passive one is used. Such modifications are important for a CubeSat design, for which the dimension and electrical power consumption must be minimized. The first modification reduces the dimensions of the setup, while the second reduces its electrical consumption. The system setup is shown in Fig. \ref{fig:ActiveSetupNano}.

A pointing subsystem is added which is necessary for aligning the CubeSat with the ground station. The setup proposed is inspired by the CubeSat Laser Infrared CrosslinK Mission (CLICK) system \cite{cahoy2019cubesat} due to its tested ability to achieve a pointing error below 1 $\mu$rad with optical data rates exceeding 20 Mbps while adhering to our \ac{SWaP} constraints. The satellite pointing system incorporates a telecom wavelength (1310 nm) from Anritsu, DFB 1310, for downlink alignment and classical data transmission purposes. The system comprises a coarse pointing stage where the satellite and the ground station align with each other using provided ephemeris information and guaranteed by the \ac{ADCS} \cite{Reves2022} and GPS. By detecting the beacon signal with a wide field of view camera, MyBlueFox from Matrix Vision, the satellite can adjust its attitude to enable the narrow field of view Quad Cell, PDQ30C from Thorlabs, to acquire the signal. This marks the initiation of the fine-pointing stage, where tracking is performed using a fast steering mirror from Mirrorcle. The option with a mirror diameter of 2.4 mm, a resonant frequency of around 860 Hz, and a maximum tilt angle of -6° to +6° should provide the necessary tracking requirements. The various optical signals are of different wavelengths and are separated or combined into the correct optical paths using dichroic mirrors from Thorlabs.

In a preliminary test, piezoelectric motored mirrors, a CMOS camera, and a closed-loop control system were used to test the satellite pointing. The camera captured laser signals at 635 and 532 nm wavelengths, an image processing algorithm determined the centroids of these signals and a PID controller ensured a swift and seamless response to pointing errors. The system tracks a dynamic reference with precision up to 3.4 mrad. The preliminary pointing and tracking control design will iterated in an upcoming free-space demonstration.

For classical data transmission, the system utilizes classical \ac{OOK}. The pointing signal is modulated and sent to the ground station.

\subsection{Optical payload design}

\subsubsection{CubeSat description and characteristics}

CubeSats are nanosatellites composed of 10 cm × 10 cm × 11.35 cm modules. Each module is referred to as 1U. For a 3U CubeSat, the components must fit a 10 x 10 x 32 cm$^3$ cuboid, have a total mass of less than 4 kilograms, and consume at most 21 Wh per orbit \cite{Neumann2018}. The 21 Wh are estimated using 30x30 cm$^2$ off-the-shelf solar panels \footnote{Values taken from \url{https://www.cubesatshop.com/wp-content/uploads/2016/07/EXA-DSA-Brochure-1.pdf}.}.

In Table \ref{tab:SWaP}, the volume (in ml), the weight (g), and the power consumption (mW), of the commercial off-the-shelf Components, are specified, (Size, weight, and Power analysis).

\begin{table}[H]
\begin{tabular}{|c|c|c|c|}
\hline
 Item   & \textbf{Volume (ml)} & \textbf{Weight (g)} & \textbf{Power (mW)}   \\ \hline \hline

 \ac{FPGA} & 110 & 94 & 500  \\
 \ac{QRNG} & 1 & - & 83  \\ 
 GPS & 49 & 45 & 500  \\
 Laser source  & 24 & 270 & 800  \\
 \ac{EOAM}   & 19 & 180 & 650  \\
 \ac{EOPM}   & 33 & 180 & 900  \\
 Variable Waveplate & 2 & 150 & 700   \\
 FC/APC Collimator & 13 & 60 & -   \\
 Passive Attenuator x 2 & 4 & 20 & -  \\ 
 Connector x 2 & 13 & 20 & -  \\ \hline
 \textbf{Alice payload} & 268 & 1019 &  4133  \\ \hline \hline

 Quad Cell & 4  & 30 &  -  \\
 Camera + Lens & 109 & 390 & 2500   \\
 Mems mirror & 3 & 30 & 85   \\
 Laser source  & 24 & 270 & 800   \\
 Dichroic Mirror x 2 & 1 & 40 &  - \\  \hline
 \textbf{Tracking payload}   & 141 & 760 & 3385   \\ \hline \hline

 Telescope  & 119 & 400 & -   \\ \hline \hline

  \textbf{Payload} &  528 & 2179 & 7518  \\ \hline \hline

 ADCS & 500  & 900 &  2000  \\
 UHF + S-band & 250 & 114 & 6000   \\
 Antennas  & 70 & 100 & 60   \\
 Batteries & 100 & 200 &  - \\ 
 Solar panels & - & 450 &  - \\  \hline
 \textbf{Platform}   & 920 & 1764 & 8060   \\ \hline \hline

 \textbf{Total} &  1448 & 3943 & 15578  \\ \hline \hline

 \textbf{3U Maximum} &  3200 & 4000  &   \\ \hline \hline

\end{tabular}
\caption{\ac{SWaP} analysis of the proposed preliminary design for a 3U CubeSat.}
\label{tab:SWaP}
\end{table}

The primary goal of the \ac{SWaP} analysis was to proactively evaluate the fit of the components within the 2U of the CubeSat. By examining their physical dimensions, volume, and relevant specifications, the goal was to determine if the components could be seamlessly integrated into the allocated space for the optical payload. This evaluation is crucial as it helps to avoid potential design iterations and modifications in the later stages of development and guides the component selection. Although these values were taken or estimated from available datasheets and may not be exact, they offer a strong basis for making informed decisions and guiding the subsequent design phases.

An estimate of the \ac{SWaP} characteristics of the system outside the payload (platform section of Table \ref{tab:SWaP}) and the telescope was also done based on similar works \cite{Neumann2018}. 

This optical system (payload) is divided into two parts, the Alice payload, and the tracking payload. 

The Alice payload will generate and encode the quantum states. These will then be sent to the telescope. As seen in Table \ref{tab:SWaP}, this subsystem's devices will only take a fraction of the total available volume. As it has most of the active components (lasers and modulators), it consumes a significant part of the power budget. Nevertheless, it only needs to be turned on during the communication window when the alignment with the ground station is guaranteed. This results in energy consumption within the mission budget.

The tracking payload houses the necessary components to guarantee a pointing error sufficiently small for the mission's success. This part of the setup occupies a bigger volume and a significant fraction of the power budget due to the use of a wide field-of-view camera but it is still below the total values available. 

\begin{figure}[H]
  \centering
  \includegraphics[width=0.7\linewidth]{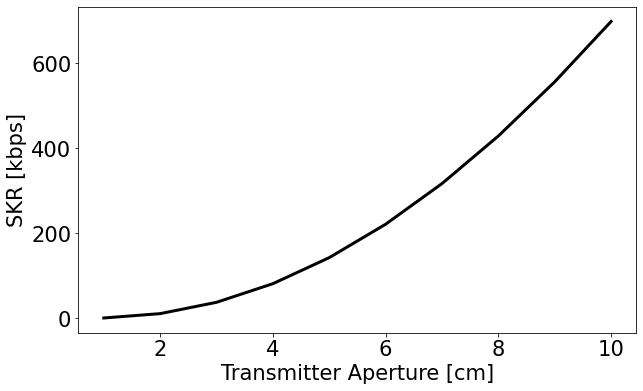}
  \caption{Optimized \ac{SKR} at zenith for different values of the transmitter aperture $D_T$, for the simplified BB84.}
  \label{fig:SKR_app}
\end{figure}

Finally, to transmit the optical signals, a telescope is necessary. To choose the aperture size for the emitter telescope, the secret key rate for the simplified BB84 as a function of the aperture was estimated, Figure \ref{fig:SKR_app}. This was done for a fixed value of the aperture of the receiver's telescope.

A 4 cm aperture is chosen to deal with the restrictions of the 3U CubeSat. For a larger CubeSat e.g. 6U or 12U, a larger aperture could be considered to increase the rates as shown in Figure \ref{fig:SKR_app}. It can be seen that for an aperture of 10 cm one can achieve a \ac{SKR} of 700 kbps.

This takes a significant part of the remaining available volume but the design is still within the limit.
The SWaP analysis with the chosen commercial off-the-shelf components demonstrates that the payload design is ready for its next stage: the custom design of optoelectronic and mechanical components, the miniaturization, and prototyping.

\subsubsection{Classical communication}

In Sidhu et al. \cite{Sidhu2022}, an estimation of classical communication cost and data storage requirements can be found.

Large satellites can work in the X and K bands, with frequencies of the order 10 - 40 GHz, which can use efficient modulations for communication rates of several Gbps \cite{article_rf_com}. Due to their size restrictions, CubeSats are much more limited with their typical bands being UHF, S, X, and Ka. The most mature bands used for CubeSat communication are VHF and UHF frequencies but there has been a shift in recent years towards S and X, with Ka being NASA’s intended band for future small satellite communications. The move to higher frequency bands has been driven by a need for higher data rates with typical numbers being around in the dozens of kbps. \cite{yost2021state}

It is possible to supplement radio communication using classical optical communication. Recently, a laser-based C2G (CubeSat-to-Ground) link from an \ac{LEO} 1.5U CubeSat at a 450 km altitude to an optical ground station was established \cite{Rose2019}. This communication link achieved a data rate of up to 100 Mbps with bit error rates near $10^{-6}$. Since, pointing and acquisition are major problems for free-space optical communications, a hybrid RF-and-optical approach is introduced in \cite{Welle2017}, where CubeSats are used as relay satellites between the \ac{GEO} satellites and the ground station using both RF and optical links.

As the system already has a laser link to the ground station through the pointing beam, it can use on-off keying to transmit information. CLICK-A, with a similar system, is expected to achieve a greater than 10 Mbps data downlink from spacecraft at an altitude of approximately 400 kilometers, to a 28-centimeter telescope on the ground \cite{9749715}. The final system would then use a hybrid RF-and-optical approach as has been shown in \cite{Welle2017}.

\section{Results}

This section showcases the results from simulations of both communication protocols in realistic scenarios, followed by a direct comparison between them. The aim is to illustrate their performance differences.

\subsection{Losses}

\begin{figure}[H]
  \centering
  \includegraphics[width=\linewidth]{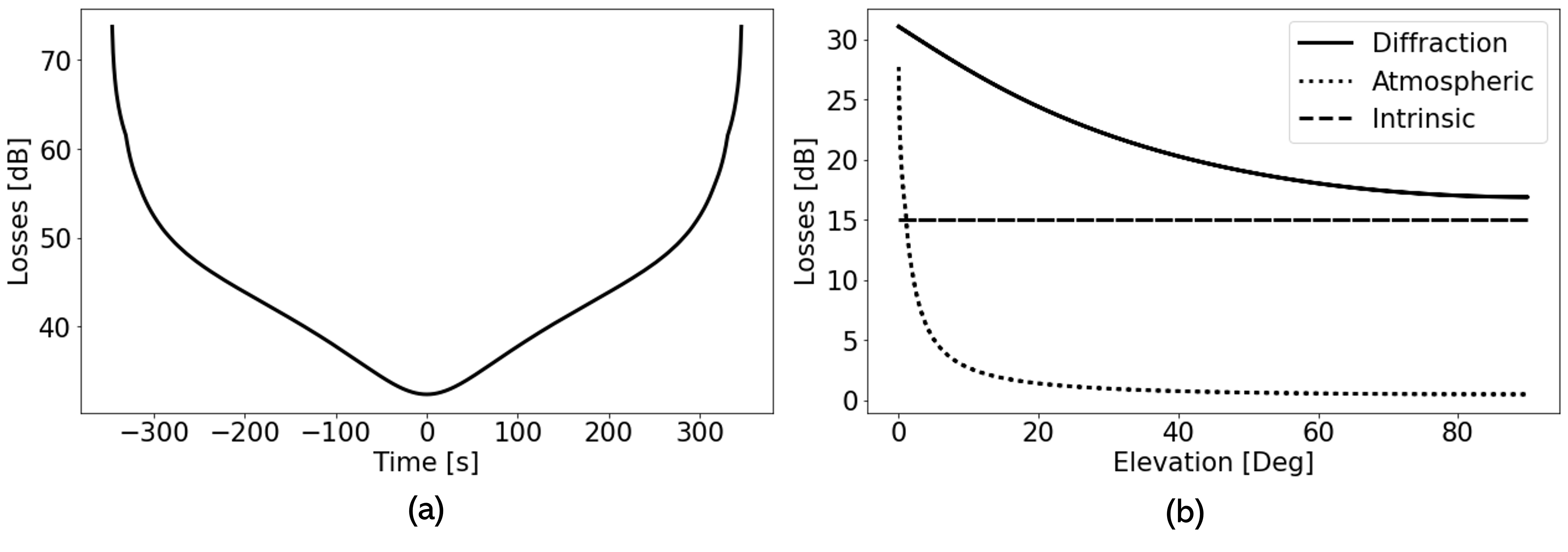}
  \caption{Estimated losses for the setup. The geometric losses are estimated using the parameters in Table \ref{tab:parameters}. A conservative value of 15 dB for the intrinsic losses is chosen. For the atmospheric losses, realistic data provided with the SatQuMA toolbox is used. (a) Optical losses as a function of time for the chosen orbit. At time t=0, the satellite is above the ground station at 90\textdegree{} of elevation (zenith). (b) Loss contributions from the different sources with respect to the elevation angle.}
  \label{fig:loss}
\end{figure}

For this analysis, three main types of losses are considered, geometric losses, atmospheric losses, and intrinsic system efficiency. Geometric losses appear from the limited receiver aperture to catch the incoming beam spread through divergence. Atmospheric losses can manifest in various forms, such as scattering, absorption, and turbulence. Intrinsic losses correspond to beam misalignment and internal losses inherent to the optical payload (e.g. optical components insertion loss and single-photon detectors efficiency). To detect the signal (single-mode), single-mode detectors are considered and although using these detectors is difficult, recent results show promising technologies with larger detection areas to overcome this challenge \cite{steinhauer2021progress}. The analysis describes the total channel's loss throughout a satellite overpass through the zenith, Figure \ref{fig:loss}.a, where the satellite's trajectory starts and ends at the horizon level (0\textdegree{} of elevation), and reaches a maximum elevation of 90\textdegree{} at $t=0$. The contribution of all types of loss for each value of the satellite's elevation is described in Figure \ref{fig:loss}.b.

The primary factor that limits the optical losses is the diffraction loss, which, throughout the trajectory, ranges from 17dB to 31 dB. At low elevations, the atmospheric loss is at the highest effect and starts to decrease exponentially with the elevation reaching 3 dB at approximately 9\textdegree{} of elevation. For the analysis of Bob's intrinsic loss, it was chosen a conservative of 15 dB.

The main factor for atmospheric losses is the transmissivity of the chosen wavelength. However, for some applications, there can be slight benefits from a different wavelength due to pollution or weather conditions. The SatQuMA toolbox provides realistic data for an 850 nm signal used in the analysis. The atmospheric losses can be evaluated also for 1550 nm using software such as MODTRAN \cite{BERK1998367} and libradtran \cite{gmd-9-1647-2016}, which is left for future work as the objective here is only to validate the design under realistic conditions, and the atmospheric transparencies for 850 nm and 1550 nm allow for transmission close to 1 Gbps using classical optical communication \cite{robinson2011overview}. As shown in \cite{maharjan2022atmospheric}, some advantages for the 1550 nm choice can be found as atmospheric turbulence has less impact, and the coherence length is longer. While these are not major advantages, they corroborate the choice to use this wavelength

\subsection{Operation Parameters}

In Table \ref{tab:parameters}, all the parameters used for the numerical simulations of the quantum communication protocols are presented.

\begin{table}[H]
\centering
\begin{tabular}{ |c|c|c| }
\hline
Parameter & Symbol & Value\\
\hline
Orbit height & $h$ & 500 km \\
\hline
Minimum Transmission Elevation & $\theta_{min}$ & 10\textdegree \\
\hline
Transmitter aperture diameter & $D_T$ & 0.04 m \\
\hline
Receiver aperture diameter & $D_R$ & 0.7 m \\
\hline
Beam waist & $\omega_0$ & 0.02 m \\
\hline
Wavelength & $\lambda$ & 1550 nm \\
\hline
Offset angle of satellite orbital plane & $\xi$ & 0 \\
\hline
Correctness parameter & $ \epsilon_C $ & $10^{-15}$ \\
\hline
Secrecy parameter & $\epsilon_S$ & $10^{-9}$ \\
\hline
Intrinsic Quantum Bit Error Rate & $\text{QBER}_I$ & 0.001 \\
\hline
Extraneous count probability & $P_\text{EC}$ & $10^{-8}$ \\
\hline
After pulse probability & $P_\text{AP}$ & 0.001 \\
\hline
Source repetition rate & $f_s$ & 1 GHz \\
\hline
\end{tabular}
\caption{Parameter values for the communication system simulations}
\label{tab:parameters}
\end{table}

The satellite will orbit in \ac{LEO} and it will be considered that no communication is possible below 10\textdegree\ of elevation, a regime where the atmospheric losses become much more important. For the quantum communication signal, a wavelength of 1550 nm is used for more efficient integration with fiber-based telecommunication networks, which in turn allows for a compact setup inside the CubeSat and the use of high-speed electro-optical modulators. The choice of parameters for the beam size and telescope apertures is done to optimize the rate while keeping the design compact enough to fit inside a 3U CubeSat.

\subsection{QKD Simulation}

For the numerical simulations, the python package SatQuMA was modified\footnote{The code is available on the Github page \url{https://github.com/QuLab-IT/QuantSatSimulator.git}.} to implement the simplified BB84, three-state and one-decoy described in \cite{Rusca2018,Rusca2018b}. SatQuMA is an open-source software that models the efficient BB84 protocol with four-state two-decoy using \acp{WCP} in a downlink configuration, described in \cite{Brougham2021,islam2022finite}. The 3-state protocol was chosen for its simpler setup, making it easier to meet the SWaP constraints. The simplified BB84 has been shown to achieve experimental secret key rates close to the ideal four-state BB84 implementation, showing that there is no performance loss by choosing this protocol \cite{Boaron_2018}. The secret key analysis for a three-state one-decoy described in was added to simulate and optimize the SKR through a satellite overpass.

The chosen orbit path transits through the zenith, ensuring maximum coverage and visibility from the ground station. In Table \ref{tab:parameters}, the values used in the simulation are given. The satellite's sun-synchronous orbit is fixed to an altitude of 500 km and the downlink transmission is made by a laser source of 1550 nm of wavelength, a common choice for high-speed optical communication networks. The telescope aperture diameter of the transmitter is fixed to 4 cm as previously explained. For the ground station telescope, an aperture of 70 cm was chosen. The beam waist is set to be half the transmitter aperture diameter, as done in SatQuMA, so as not to clip too much of the Gaussian beam. This choice affects the system's performance, as a larger beam waist would lead to better signal strength and more efficient transmission. 

\begin{figure}[H]
  \centering
  \includegraphics[width=0.7\linewidth]{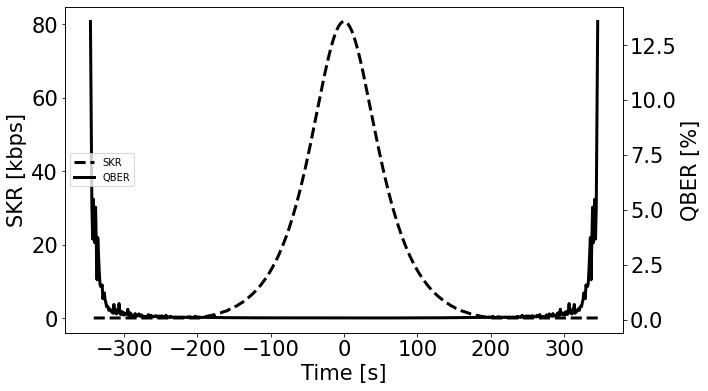}
  \caption{Satellite-to-ground \ac{QKD} Simulation. Secret Key Rate (dashed line) and Quantum Bit Error Rate (solid line) over a satellite pass.}
  \label{fig:simul}
\end{figure}

To optimize the performance of the satellite to ground station communication system, the parameter $P^{B}_{Z}$ was fixed to $0.9$, which is a common value for a BS, and the parameters $k$, $P_{k}$ and $P^{A}_{Z}$ (with $k = \{\mu_1,\mu_2\}$) were set to vary according to the losses of the system and the transmission time window. Figure \ref{fig:simul} shows the numerical simulation of optimized \ac{SKR} and \ac{QBER} during a satellite overpass.

The simulation assumes a perfect satellite overpass with a maximum elevation of 90 degrees. In this analysis, the \ac{SKR} values range up to 80.8 kHz, and the total transmission window is approximately 304 seconds per pass. Consequently, the total secure block size after one satellite pass is approximately 9.9 Mbits. Each value of \ac{SKR} was obtained by optimizing the \ac{SKL} within a 1-second time window (time interval between values of channel's attenuation, see Figure \ref{fig:loss}.a). The Secret Key Length encompasses both the transmitted secret key bits and the final leaked bits, denoted as $\lambda_{EC}$, used for \ac{QBER} deduction.

The minimum \ac{QBER} value occurs at the zenith, at 0.08\%, and increases rapidly for lower elevations. The simulation is designed to maximize the \ac{SKR}, which results in the optimizer being unable to converge at a fixed value for \ac{QBER}, when the \ac{SKR} is zero, as evidenced by the oscillations in the figure. Nevertheless, within the total transmission window, the \ac{QBER} remains below 1\%.

\begin{figure}[H]
  \centering
  \includegraphics[width=\linewidth]{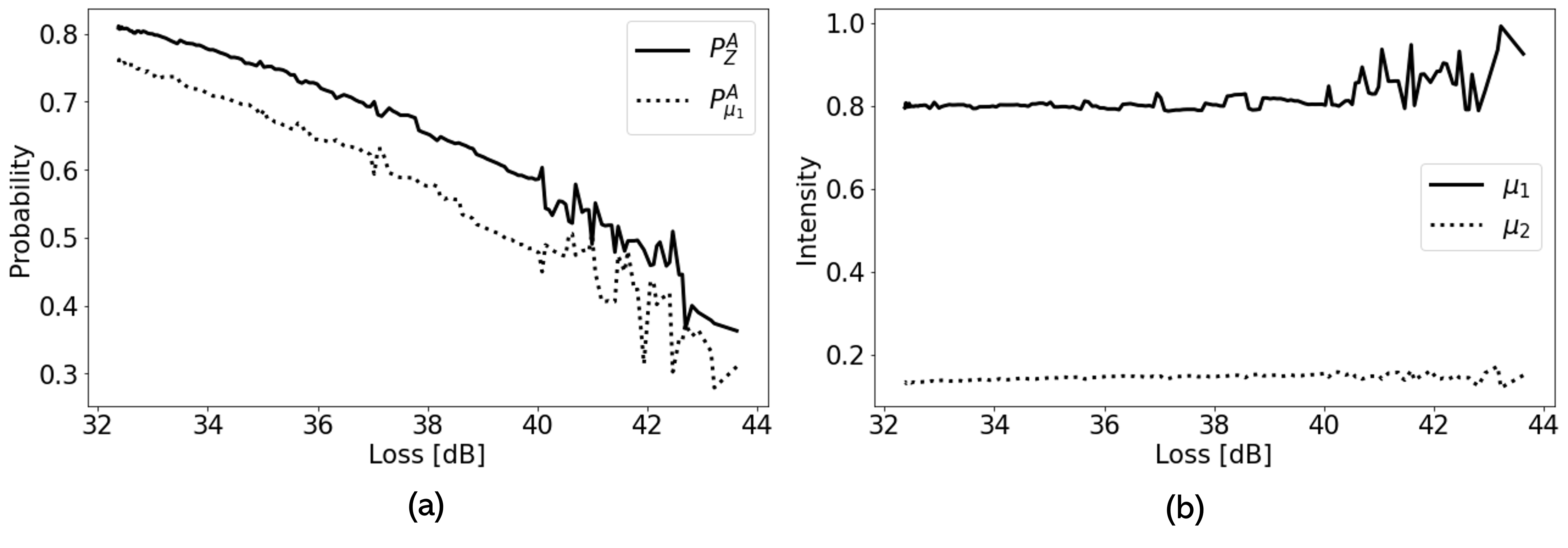}
  \caption{Optimised parameters as a function of the system loss. (a) Probabilities of Alice sending an $Z$ basis state, $P_{Z}^{A}$ (solid line), and of sending a state state, $P_{\mu_1}$ (pointed line). (b) Intensity of state signal $\mu_1$ (solid line) and decoy signal $\mu_2$ (pointed line)}
  \label{fig:parameters}
\end{figure}

Figure \ref{fig:parameters} represents the optimal set of values for the protocol parameters as a function of the total loss of the system. To ensure a maximal \ac{SKR}, the values of probabilities of $P^{A}_{Z}$ and $P_{\mu_1} $ decrease rapidly with the increase of loss, the values of the intensities $\mu_1$ and $\mu_2$ vary very little compared with the probabilities but their value increase slightly with the system losses. For high values of loss (close to a zero \ac{SKR}), the simulator has difficulty converging to a set of parameters. However, there is a clear tendency in the figures. After 43.6 dB of loss, the system cannot maintain transmission of secret key bits. Thus, the values of the parameters can no longer optimize the communication. The optimal values for the parameters at the zenith position are presented in Table \ref{tab:optparam}.

\begin{table}[H]
\centering
\begin{tabular}{ |c|c|c| }
\hline
Parameter & Symbol & Value\\
\hline
Intensity 1 & $\mu_1$ & 0.81 \\
\hline
Probability of sending intensity 1 & $P_{\mu_1}$ & 0.76 \\
\hline
Intensity 2 & $\mu_2$ & 0.12 \\
\hline
Probability of sending intensity 2 & $P_{\mu_2}$ & 0.24 \\
\hline
Probability Alice sends an $Z$ basis signal & $P^{A}_{Z}$ & 0.88 \\
\hline
Probability Bob measures an $Z$ basis signal & $P^{B}_{Z}$ & 0.9 \\
\hline
\end{tabular}
\caption{Optimal Communication parameter values for the zenith}
\label{tab:optparam}
\end{table}

\subsection{QKPC Simulation}

The \ac{QKPC} security arguments to a realistic channel are applied, using the same data used for the \ac{QKD} simulations. The number of photons detected by Bob is $\eta\mu$, where $\eta$ are the same losses considered in the \ac{QKD} simulations and shown in Fig. \ref{fig:loss}.a.

The number of photons detected by Eve is $\gamma\eta\mu$. The realistic value of $\gamma = 0.1$ is chosen based on \cite{Vazquez2021}.

\begin{figure}[H]
  \centering
  \includegraphics[width=\linewidth]{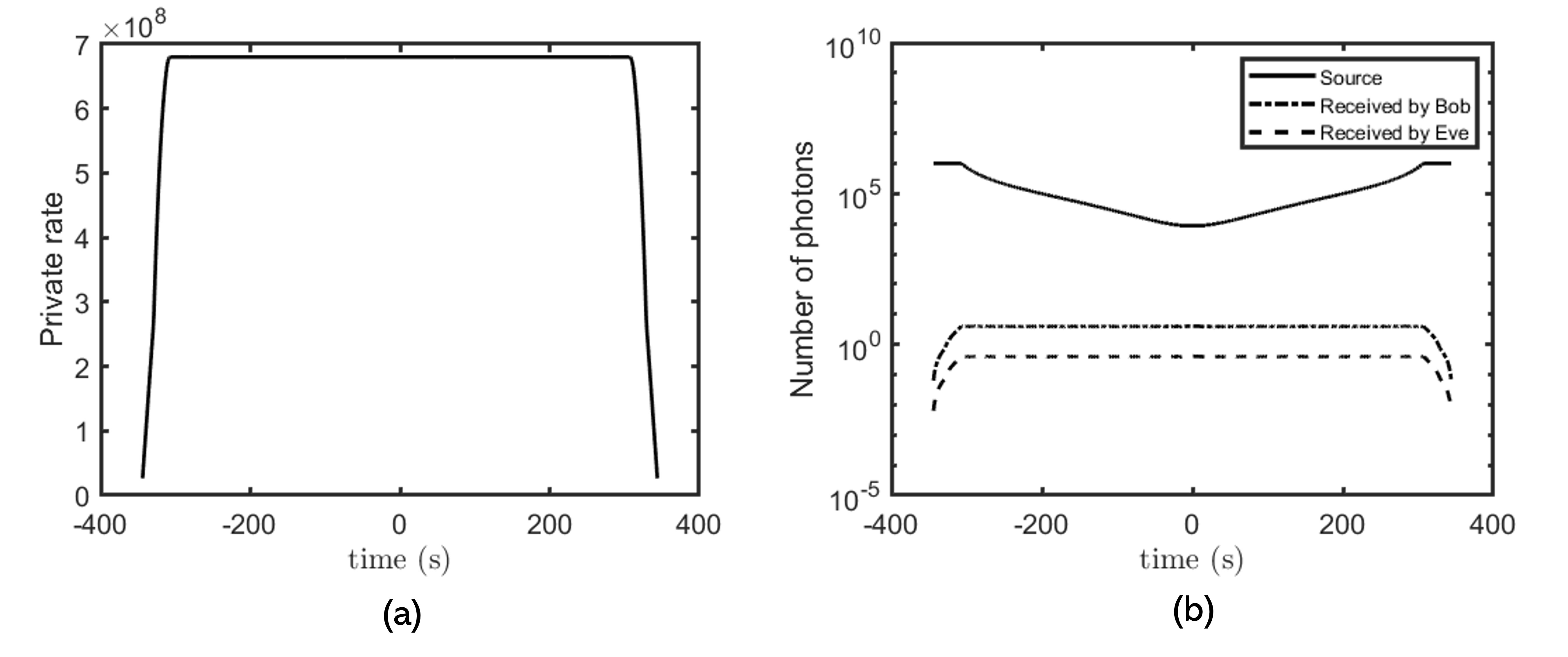}
  \caption{a) \ac{QKPC} rates based on the private capacity. b) Number of photons required to maximize the private capacity versus time.}
  \label{fig:QKPCresults}
\end{figure}

In Fig. \ref{fig:QKPCresults}, the results of the simulations are shown. In Fig. \ref{fig:QKPCresults}.a, the rate versus elevation is presented. It is seen for a wide range of elevations that, \ac{QKPC} can provide a secret transmission rate of 700 MHz. In Fig. \ref{fig:QKPCresults}.b, it can be seen how the number of photons must be varied in order to guarantee the optimal transmission rate. It was found that for optimal rates, the source must send about one million photons per pulse, Bob receives 3 - 4 photons per pulse, while Eve only receives about 0.3 - 0.4 photons per pulse. This ensures that while Bob can efficiently discriminate the coherent state from the vacuum, for Eve it is much more difficult.

Figure \ref{fig:QKPCresults} also shows that the communication window is wider for the \ac{QKPC}. In the \ac{QKD} protocol, the losses have to drop below a certain value for the communication to start. In our simulations, only during approximately 304 seconds in a pass will the \ac{QKD} rate be strictly positive. For the \ac{QKPC} protocol, the system can adapt to the losses by varying the number of photons it sends. Therefore, in a satellite pass, it can communicate as soon as there is a line-of-sight with the receiver, extending the communication window close to the time of a pass (around 600 seconds) while maintaining the optimal secret transmission rate of 700 MHz. This means the total number of secret bits sent (or generated) in a pass can be close to 420 Gbits as opposed to the 9.9 Mbits of the \ac{QKD} protocol. Since the \ac{QKPC} protocol can sustain higher losses, it can also work during situations where the \ac{QKD} cannot operate. These situations include bad weather conditions and daytime. Nevertheless, it is important to note that these protocols serve different purposes. \ac{QKD} is used for key exchange and provides unconditional security while the \ac{QKPC} is used to transmit a direct message and offers security under more relaxed assumptions.

\section{Conclusion}

Quantum communication in space is a very promising research field in what concerns information privacy. Recent efforts have focused on \ac{QKD} solutions, nevertheless, that particular class of quantum communication protocols is far from achieving practical rates for telecommunications. This work goes beyond such approaches, by introducing a nanosatellite design capable of performing both \ac{QKD} and another class of protocols called \ac{QKPC}. In this article, a preliminary design for a 3U CubeSat quantum communications downlink is proposed. The CubeSat serves as a platform to implement various quantum communication protocols. This versatility is demonstrated with two examples: \ac{QKD} with the simplified BB84 and \ac{QKPC}. 

The design is validated via a \ac{SWaP} analysis using commercial off-the-shelf components. It is argued that all the mission requirements, including pointing and classical communication, can be achieved in a 3U CubeSat. The feasibility of \ac{LEO} communication is shown using the design via numerical simulations of the simplified BB84 and \ac{QKPC}. In the case of \ac{QKD}, we expand an existing toolbox called SatQuMA to achieve a realistic simulation of the simplified BB84 in a downlink configuration. It is found that, under realistic conditions, at zenith, a \ac{SKR} is obtained for the simplified BB84 slightly over $80$ kHz and a \ac{QBER} slightly larger than $0.07\ \%$. It is shown the \ac{QKPC} scheme achieves an optimal 700 MHz private communication rate for a wide range of elevations, in fact during most of the communication time.

Future study directions to validate the solution include building a demonstration setup with portable optical breadboards, one for Alice, and one for Bob, and building a prototype of the CubeSat which can be used for space validation. There are several options to further miniaturize the solution, and the optimal solution will most likely involve integrated optics. Hence, another important direction to follow is to design photonic integrated circuits implementing at least part of the optical payload proposed. For example, the generation of weak coherent pulses for \ac{QKD} and \ac{QKPC} can be done on a photonic integrated circuit, and other CubeSat missions with quantum communication payloads have already started investigating/using those \cite{zhu2022experimental, de2022satellite}. Other future work planned will include the compatibility of the CubeSat system with the CCSDS standards and with current/planned ground stations. Naturally, such a miniaturization of the optical payload will allow for a better performance of the CubeSat, for example in the most limiting properties such as pointing and classical communication.

Regarding applications, besides long-distance quantum communication, this solution could serve as a payload for free-space quantum communication using airplanes or drones.

\section*{List of Abbreviations}

\begin{acronym}[H.264/SVC]

    \acro{ADCS}{Altitude Determination and Control System}
    \acro{DFB}{Distributed-Feedback}
    \acro{EOAM}{Electrooptic Amplitude Modulator}
    \acro{EOPM}{Electrooptic Polarization Modulator}
    \acro{FPGA}{Field-Programmable Gate Array}
    \acro{GEO}{Geostacionary Orbit}
    \acro{LEO}{Low-Earth Orbit}
    \acro{MEO}{Medium-Earth Orbit}
    \acro{OOK}{On-Off Keying}
    \acro{PC}{Polarization Controller}
    \acro{QBER}{Quantum Bit Error Rate}
    \acro{QKPC}{Quantum Keyless Private Communication}
    \acro{QKD}{Quantum Key Distribution}
    \acro{QRNG}{Quantum Random Number Generator}
    \acro{SKL}{Secret Key Length}
    \acro{SKR}{Secret Key Rate}
    \acro{SWaP}{Size, Weight and Power}
    \acro{WCP}{Weak Coherent Pulse}
    \acrodefplural{WCP}{Weak Coherent Pulses}

\end{acronym}

\section*{Declarations}

\textbf{Availability of data and materials}

The repository \url{https://github.com/QuLab-IT/QuantSatSimulator.git} contains the Python code developed in this work. 

\textbf{Competing interests}

The authors declare no competing interests.

\textbf{Funding}

The authors thank the support from Instituto de Telecomunicações, namely through project QuantSat-PT (UIDB/50008/2020) and the support from the European Commission (EC) through project PTQCI (DIGITAL-2021-QCI-01).
E.Z.C. acknowledges funding by FCT/MCTES - Fundação para a Ciência e a Tecnologia (Portugal) - through national funds and when applicable co-funding by EU funds under the project UIDB/50008/2020. E.Z.C. also acknowledges funding by FCT through project 2021.03707.CEECIND/CP1653/CT0002. D.R. thanks the Galician Regional Government (consolidation of Research Units: AtlantTIC), MICIN with funding from the European Union NextGenerationEU (PRTR-C17.I1) and the Galician Regional Government with own funding through the “Planes Complementarios de I+D+I con las Comunidades Autónomas” in Quantum Communication and The European Union’s Horizon Europe Framework Programme under the project “Quantum Security Networks Partnership” (QSNP, grant agreement No 101114043).

\textbf{Author's contributions}

P.N.M. and G.L.T. worked on the setup design, numerical simulations, and writing the manuscript. D.P. contributed to discussions on the satellite design. E.Z.C. supervised the work and contributed to every step of it. R.R., P.A., M.N., R.F. ad D.R. co-supervised the work.

\textbf{Acknowledgments}

Not applicable.

\bibliography{bibl}
\bibliographystyle{unsrt}

\end{document}